\documentstyle[preprint,revtex]{aps}
\begin{document}
\begin{title}
Exact result on the Mott transition in a two-dimensional model\\ of strongly
correlated electrons
\end{title}
\vskip0.5truecm
\author {Fr\'ed\'eric Mila}
\vskip0.5truecm
\begin{instit}
Laboratoire de Physique Quantique\\
Universit\'e Paul Sabatier\\
31062 Toulouse Cedex\\
France
\end{instit}

\begin{abstract}
We study the properties of a quarter-filled
system of electrons
on a square lattice interacting through a local repulsion $U$ and a
nearest-neighbour repulsion
$V$ in the limit $V=+\infty$. We identify the ground-state for $U$ large enough
and
show that domain walls appear below a critical value $U_c=4t/\pi$. We argue
that
this corresponds to a metal-insulator transition due to Mott localization.
\end{abstract}
\vskip 1.truein

\vskip 0.1 truein
\newpage
A lot of attention is currently devoted to the
problem of metal-insulator transitions (MIT) driven by correlations in
electronic
systems\cite{mott}. While this problem has some direct experimental relevance,
the main
reason for this renewed interest lies in the possibility that, on the
metallic side of the transition, but close enough to the boundary, the electron
gas might no
longer behave as a Fermi liquid. The model most often considered to study this
phenomenon is the Hubbard
model defined by
\begin{equation}
H=-t\sum_{<ij>,\sigma}(c{}^\dagger_{i\sigma}c{}^{\ }_{j\sigma}+h.c.)
+U\sum_i n_{i\uparrow}n_{i\downarrow}
\label{1}
\end{equation}
\indent where the symbol $<ij>$ means summation over pairs of nearest
neighbours. At half-filling, this model is believed to undergo a transition
from a metallic phase to an insulating one as $U$ increases with a critical
value of the
repuslion $U_c$ that
depends on the lattice and on the dimension. The main source of concern in
using
this model to study Mott transitions
is that the MIT that occurs is often due to the
appearance of a
spin-density wave (SDW) and not to Mott localisation as first pointed out by
Slater\cite{slater}. This is in particular true for
lattices that lead to perfect nesting, like e.g. the cubic lattice, in which
case
$U_c=0$. This was
confirmed by the Bethe ansatz solution of the one-dimensional Hubbard model by
Lieb and Wu\cite{lieb}. On the contrary, Mott localisation is expected to
yield a finite value for $U_c$.
In that respect, to identify models that exhibit a MIT
due to Mott
localisation is an important issue.

In this paper, we concentrate on the extended Hubbard model defined by
\begin{equation}
H=-t\sum_{<ij>,\sigma}(c{}^\dagger_{i\sigma}c{}^{\ }_{j\sigma}+h.c.)
+U\sum_i n_{i\uparrow}n_{i\downarrow}
+V\sum_{<ij>} n_in_j
\label{2}
\end{equation}
\indent on a square lattice at
quarter-filling. The central motivation to study such an extension of the
Hubbard model comes from the one-dimensional case which is known to have a
MIT when $U$ and $V$ increase. The first result on that problem was obtained
about 20 years ago by Ovchinnikov\cite{ovchinnikov} who showed that when
$U=+\infty$ there is a
MIT for $V=2t$. Quite recently, an extensive analysis of the 1D model for
arbitrary $U$ and $V$ has been carried on\cite{mila}, and it has been shown
that the MIT
occurs along a line in the ($U$,$V$) plane going from ($+\infty$,$2t$), which
is
nothing but Ovchinnikov's result, to
($4t$,$+\infty$). The shape of the line was obtained numerically, but it
was possible to prove analytically that the point $U=4t$, $V=+\infty$ belongs
to
the boundary.
That this transition is of the Mott type is quite
clear for $V=+\infty$ as
there is a full spin degeneracy and hence no SDW in the
insulating phase. In dimension 2, we expect to get
rid of the SDW-driven transition because this model is not half-filled but
quarter-filled, in which case there is no perfect nesting for the square
lattice.

The essential reason that allowed an exact determination of $U_c$ for
$V=+\infty$ in the one-dimensional case is in fact independent of the
dimension and
can be summarized as follows. A convenient way to take the fact that
$V$ is infinite into account is to include it as a constraint that restricts
the Hilbert space to states with no pair of particles sitting next to
eachother. Then, it is easy to see that,
for $U$ large enough, the ground state is the checkerboard
state
obtained by putting one particle on each site of one sublattice and no particle
on the other sublattice (see fig.1). This is an eigenstate because all the
states that could
be connected to it by the hopping term are not in the Hilbert space due to the
constraint imposed by $V=+\infty$. Besides all the other states have at least
one doubly occupied site
and are certainly higher in energy when $U$ is large enough. The crucial point
is that this checkerboard state is always an eigenstate regardless of the
value of $U$ because of the constraint. However, when $U$ decreases, it is no
longer so costly to make local pairs, and states having such local pairs can be
lower in energy than the checkerboard state due to a gain of kinetic energy.

There is an important difference however between the one-dimensional and
two-dimensional cases. In 1D, a single vacancy is already a mobile defect:
It splits into two doublets that can move away from each other\cite{fowler}. In
2D, any
defect made of a {\it finite} number of vacancies cannot move far
away. To see
this, consider a rectangle that encloses all the vacancies and whose sides go
through empty sites of the checkerboard configuration (see fig.2). It is easy
to see
that all the particles located outside this rectangle are not able to move. Due
to the constaint $V=+\infty$, a particle is able to move to an empty site if
all the other neighbouring sites of that empty site are also empty. But the
empty
sites available to the particles outside the rectangle are either outside the
rectangle or on its boundary, and they have at least two particles as
neighbours.

So the only chance to get a metallic state, that is charge propagation, is to
consider defects with an infinite number of vacancies. From what we just saw,
these vacancies have to constitute a connected set in order to allow for charge
propagation. The cheapest way in terms of the number of vacancies per unit
length is, like in a ferromagnet, to make a domain wall between two
checkerboard
configurations translated from each other by one lattice spacing (see fig.3).
Due to the constraint, one has to remove a row after performing the
translation.
So the density of vacancies per unit length in such a defect is equal to the
density of particles in a row, that is  1/2. To create a vacancy, one has to
create a local pair somewhere else with an energy cost $U$. So the potential
energy per unit length of such a domain wall is $U/2$.

Now, such a defect is able to gain kinetic energy. The elementary move is
depicted in
fig.3. Let us keep track of the position of the domain wall by drawing a line
that joins the middles of two neighbouring empty sites. Let us also assume
for the moment that one point is fixed. Taking that point as a reference, the
position of the wall is determined by the succession of bonds going up or down.
A move is possible only if two sucessive bonds are going in opposite ways, and
it consists of exchanging them. If we represent a configuration by a series of
0
and 1, where 0 stands for a down-going bond and 1 for an up-going bond, a
configuration on a periodic finite system has as many 0 as 1, and the
elementary
moves are 01 $\rightarrow$ 10 and {\it vice versa}, with amplitude $-t$. This
is equivalent to the
eigenvalue problem of spinless fermions on a 1D chain at half-filling with
amplitude $-t$. The kinetic energy per unit length is thus given by
\begin{equation}
{1 \over 2\pi} \int_{-\pi/2}^{\pi/2} -2t\cos k \ \  dk = {-2t\over \pi}
\end{equation}
\indent This mapping relies on the boundary condition that one point of the
wall
is fixed. In the limit of infinite systems, this boundary condition plays
no role, and the kinetic energy of a domain wall is effectively equal to
$-2t/\pi$. The total energy per unit length is then given by $U/2-2t/\pi$. It
vanishes at a critical value $U_c=4t/\pi$. Below that value, domain walls are
present in the ground-state.

The presence of domain walls in the ground-state does not guarantee by itself
that the state is metallic. For instance, in the case of the
weak-coupling Hubbard model, it
is believed that the domain walls that appear away from half-filling
crystallize, and that the system remains insulating up to a certain doping
\cite{schu}. However, this effect is due to the large extension of the domain
walls which gives rise to a long-range interaction between them. In the present
case, the interaction between walls is  local, and such a
crystallization is unlikely to
occur. So we believe that the appearance of domain walls in
the ground-state corresponds to a metal-insultor transition. A more careful
analysis of the interaction between domain walls is needed however to settle
that issue.
Concerning the nature of the transition, we note that, as in the 1D case, there
is a full spin degeneracy in the insulating state, and hence no SDW. Thus the
MIT cannot be due to a SDW and has to come from Mott localization.
However, the insulating ground-state is a commensurate charge-density wave, so
that the lattice translation symmetry is broken, in agreement with a general
argument of Lee and Shankar\cite{lee}.

Finally, let us comment on the possible applications of these results.
This paper provides an example of Mott transition in dimension 2 for
which the critical value of the interaction is known exactly. This might be
useful to check approximate
methods used in the context of
Mott transitions. If one can extend the results to finite values of $V$, this
model might also provide an example of Mott transition for reasonable values of
$U$ and $V$\cite{note}. In that respect, we should note that similar results
can be
obtained for bosons. Charging effects in Josephson junction networks would then
be a good realization of these ideas.

I acknowledge useful discussions with K. Penc, D. Poilblanc, S. Sorella, H.
Tsunetsugu, T. Ziman and X.
Zotos.

\figure{Checkerboard configuration. Crosses stand for electrons with spin up
or down; small circles stand for empty sites.}

\figure{Example of a defect made of a few vacancies (heavy circles).}

\figure{Typical configurations of a domain wall. The dashed line corresponds to
the new configuration obtained after an elementary move from the original one
(solid line).}


\begin{references}

\bibitem{mott} The original idea can be found in N. F. Mott, Phil Mag {\bf 6},
287 (1961). Among the recent developements, let us mention the work in infinite
dimension (see e.g. M. J. Rozenberg et al, Phys Rev Lett {\bf 69}, 1236 (1992),
A. Georges and W. Krauth, Phys Rev Lett {\bf 69}, 1240 (1992), V. Janis and D.
Vollhardt, Int Journ Mod Phys B {\bf 6}, 731 (1992) and M. Jarrell,
Phys Rev Lett {\bf 69}, 168 (1992)). Another recent development of the field
can
be found in M. Imada, J Phys Soc Japan {\bf 62}, 1105 (1993). References to
earlier work can be found in these papers.

\bibitem{slater} J. C. Slater, Phys Rev {\bf 82}, 538 (1951).

\bibitem{lieb} E. H. Lieb and F. Y. Wu, Phys. Rev. Lett. {\bf 20}, 1445 (1968)

\bibitem{ovchinnikov} A. A. Ovchinnikov, Sov Phys JETP {\bf 37}, 176 (1973).

\bibitem{mila} F. Mila and X. Zotos, Europhys Lett {\bf 24}, 133 (1993);
K. Penc and F. Mila, Phys. Rev. B, in press.

\bibitem{fowler} M. Fowler and M. W. Puga, Phys. Rev. B {\bf 18}, 1 (1978).

\bibitem{schu} D. Poilblanc and T. M. Rice, Phys. Rev. B. {\bf 39}, 9749
(1989);
H. Schulz, Phys. Rev. Lett. {\bf 64}, 1445 (1990).

\bibitem{lee} D. H. Lee and R. Shankar, Phys. Rev. Lett. {\bf 65}, 1490 (1990).

\bibitem{note} If $V$ is large but finite, a magnetic coupling $J_{eff}$ of
order $t^4/UV^2$ will appear due to superexchange, leading to antiferromagnetic
order. However, this energy scale is much smaller than the charge gap, which is
of order $U$, so that the insulating state can still be considered as a Mott
insulator.
\end{references}
\end{document}